\documentclass[reprint,aps,prl,showpacs]{revtex4-1}
\usepackage{natbib}
\usepackage{amssymb,amsmath,graphicx,mathtools}
\begin{document}

\title{Ion-Photon Entanglement and Bell Inequality Violation with \Ba}
\author{Carolyn Auchter}
\email{cauchter@uw.edu}
\homepage{http://depts.washington.edu/qcomp/}
\author{Chen-Kuan Chou}
\author{Thomas W. Noel}
\author{Boris B. Blinov}
\affiliation{Department of Physics, University of Washington, Seattle, WA 98195}
\newcommand{\Ba}{\textsuperscript{138}Ba\textsuperscript{+} }

\begin{abstract}  We report on the demonstration of ion-photon entanglement and Bell inequality violation in a system of trapped \Ba ions.  Entanglement between the Zeeman sublevels of the ground state of a single \Ba ion and the polarization state of a single 493 nm photon emitted by the ion with a fidelity of $0.84\pm0.01$ was achieved, along with a Bell signal of 2.3, exceeding the classical limit of 2 by over eight standard deviations.  This system is a promising candidate for a loophole-free Bell inequality violation test as the wavelengths of the transitions of \Ba are in the visible region and thus suitable for long range transmission over fiber optic cable.
\end{abstract}
\received{\today}

\maketitle
\section{Introduction}

\paragraph*{}  The generation of entanglement is an essential tool for the realization of scalable quantum computing and long distance quantum communication.  The ability to entangle a photon for reliable long range quantum communication and another particle with a long-term quantum memory would allow for the construction of such a scalable quantum network.  Large strides have been made towards this in many different physical media.  Entanglement and the means for communication have been demonstrated most recently in quantum dots \cite{DeGreve13, Gao12, Juska13}, nitrogen vacancy centers in diamond \cite{Togan10, Neumann08, Bernien12}, neutral atoms \cite{Wilk10, Rosenfeld08}, atomic ensembles \cite{Li13, Chou05}, superconducting qubits \cite{Eichler12, Steffen06, Berkley03}, and ions \cite{Moehring07, Blinov04}. Here we report on entanglement and Bell inequality violation in a system of a photon and a \Ba ion.  This system is particularly well suited for quantum computation \cite{Monroe12} and communication because its transitions have relatively long wavelengths making them more suitable for long range fiber optic transmission.

\paragraph*{}  In this experiment, entanglement was produced by spontaneous emission of a photon from an excited \Ba ion with multiple decay channels similar to \cite{Blinov04}.  The two entangled qubits were the Zeeman sublevels of the $6S_{1/2}$ ground state of the ion and the emitted photon's polarization.  Entanglement was verified by performing state detection on each qubit in two non-orthogonal bases.  

\section{Theoretical Background}

\paragraph*{}  The famous 1935 paper by Einstein, Podolsky, and Rosen concluded that the description of reality given by the wave function in quantum mechanics was not complete \cite{Einstein35}.  This was based on the observation that entangled quantum systems require nature to be nonlocal or non-real.  However in 1964, John Bell demonstrated that the principles of locality and realism assumed by Einstein, Podolsky, and Rosen in their paper were inconsistent with the predictions of quantum mechanics \cite{Bell64}.  The mathematical formalism he presented gave an inequality that must be obeyed by any theory satisfying local realism.  Many experimental tests have shown violations of some form of the Bell inequality \cite{Giustina13, Matsukevich08, Waldherr11, Ansmann09, Hasegawa03, Walther05}, however none have simultaneously closed both of the major loopholes (the detection, or ``fair sampling," loophole and the locality loophole) associated with such a measurement.

\paragraph*{}  In our experiment, we demonstrate violation of the Bell inequality of the form suggested by Clauser, Horne, Shimony, and Holt (CHSH) \cite{Clauser69}.  Experimental tests of the CHSH inequality have three requirements: (1) repeated creation of entangled pairs of qubits, (2) the ability to rotate each qubit independently through a polar angle $\theta_a$, $\theta_b$ on the Bloch sphere, and (3) state measurement of each qubit.  CHSH show that all theories satisfying local realism must obey the following inequality:
\begin{equation} 
S = |P(\theta_{a}, \theta_{b}) - P(\theta_{a}, \theta_{b'})| + |P(\theta_{a'}, \theta_{b}) + P(\theta_{a'}, \theta_{b'})| \leq 2
\label{CHSH}
\end{equation}
where
\begin{equation}
\begin{split}
P(\theta_{a}, \theta_{b})&= f_{00}(\theta_{a}, \theta_{b}) + f_{11}(\theta_{a}, \theta_{b})\\&\quad - f_{01}(\theta_{a}, \theta_{b}) - f_{10}(\theta_{a}, \theta_{b})
\label{CorrFunc}
\end{split}
\end{equation}
is a correlation function measurement, and ${f_{\alpha\beta}(\theta_a, \theta_b)}$ is the fraction of the total events in which qubits A and B are found to be in states denoted by $\alpha$ and $\beta$ respectively following rotations of the qubits by polar angles $\theta_a$ and $\theta_b$ respectively on the Bloch sphere.

\paragraph*{}  Quantum mechanics predicts that the CHSH inequality can be violated for particular entangled states and measurements.  For example, given the Bell state ${|\Phi^+\rangle = \frac{1}{\sqrt{2}}(|0\rangle_A|0\rangle_B + |1\rangle_A|1\rangle_B)}$, one calculates the corresponding correlation function ${P(\theta_a, \theta_b) = \cos(\theta_a - \theta_b)}$.  This results in a violation of Eq. (\ref{CHSH}) for certain angles and maximum violation $S = 2\sqrt{2}$ for $\theta_{a,a'} = 0$, $\pi/2$ and $\theta_{b,b'} = \pi/4$, $3\pi/4$.

\paragraph*{}  The Bell signal measured in an experiment can be reduced, however, by infidelity in the state that is produced.  A measure of entanglement fidelity can be found by calculating a state's overlap with the appropriate maximally entangled Bell state \cite{Bennett96}.  Thus with respect to the particular Bell state ${|\Phi^+\rangle = \frac{1}{\sqrt{2}}(|0\rangle_A|0\rangle_B + |1\rangle_A|1\rangle_B)}$, the fidelity is given by the expectation value of the density matrix ($\rho$) in the Bell state:
\begin{equation}
\begin{split}
F = \langle\Phi^+|\rho|\Phi^+\rangle= \frac{1}{2}(\rho_{11} + \rho_{44} + \rho_{14} + \rho_{41})
\label{Fidelity1}
\end{split}
\end{equation}
where $\rho$ is given in the computational basis, $\{|00\rangle, |01\rangle, |10\rangle, |11\rangle\}$.  The first two density matrix elements are just the correlation probabilities of detecting the states $|0\rangle_A$ with $|0\rangle_B$ and $|1\rangle_A$ with $|1\rangle_B$.  These correlation probabilities ($\rho_{ii}$) are given by the probability of detecting qubit $B$ in the state corresponding to $i$ multiplied by the conditional probability of detecting qubit $A$ in the state corresponding to $i$ given that the state of qubit $B$ is the state corresponding to $i$.  The second two elements can be determined by rotating each qubit's measurement basis by the polar angle $\frac{\pi}{2}$ on the Bloch sphere.  Then ${\rho' = R_{\frac{\pi}{2}}(\phi) \rho R^{\dagger}_{\frac{\pi}{2}}(\phi)}$, where $R_{\frac{\pi}{2}}(\phi)$ is the $\frac{\pi}{2}$ polar rotation operator with relative phase $\phi$.  With $\phi$ set to zero, one finds that 
\begin{equation}
\begin{split}
\rho_{14} + \rho_{41} &= \rho'_{11} + \rho'_{44} - \rho'_{22} - \rho'_{33} - \rho_{23} - \rho_{32} 
\\&\geq \rho'_{11} + \rho'_{44} - \rho'_{22} - \rho'_{33} - 2\sqrt{\rho_{22}\rho_{33}}
\label{density}
\end{split}
\end{equation}
Thus we find a lower bound for the fidelity expressed in terms of the correlation probabilities in the original and rotated bases:
\begin{equation}
\begin{split}
F \geq \frac{1}{2}(\rho_{11} + \rho_{44} - 2\sqrt{\rho_{22}\rho_{33}} + \rho'_{11} + \rho'_{44} - \rho'_{22} - \rho'_{33})
\label{Fidelity2}
\end{split}
\end{equation}
One also finds the expected Bell signal for the angles of maximal violation from the density matrix in the unrotated and rotated bases (with the relative phase of the rotation $\phi=0$) of:
\begin{equation}
\begin{split}
S = \sqrt{2}(\rho_{11} + \rho_{44} - \rho_{22} - \rho_{33} + \rho'_{11} + \rho'_{44} - \rho'_{22} - \rho'_{33})
\label{Bellsignal}
\end{split}
\end{equation}

\section{Experimental Setup}

\paragraph*{}  The ion was confined in a linear Paul trap and Doppler cooled on the $6S_{1/2}$ to $6P_{1/2}$ transition with a 493 nm laser.  A 650 nm laser was used to repump from the long-lived $5D_{3/2}$ state (lifetime $\approx 80$ s \cite{Gurell07}) to which the $6P_{1/2}$ state may decay.  An applied magnetic field $B \approx 2.4$ gauss provided a quantization axis, and a circularly polarized beam of 493 nm light aligned parallel to the quantization axis was used for optical pumping to either of the two Zeeman sublevels of the $6S_{1/2}$ ground state.

\paragraph*{}  Ionic qubit state detection was accomplished by the use of a stabilized 1762 nm fiber laser to drive the ``shelving" transition from the $6S_{1/2}$ ($m_J = -\frac{1}{2}$) ground state sublevel (defined as $|\downarrow>$) to the $5D_{5/2}$ ($m_J = -\frac{5}{2}$) metastable sublevel (lifetime $\approx 30$ s) using adiabatic rapid passage sweeps \cite{Noel12, Wunderlich07}.  Once excited to the $5D_{5/2}$ shelved state, the ion is removed from the cooling cycle and will appear ``dark" while the cooling lasers are incident on the ion, whereas an unshelved ion will be ``bright."

\begin{figure}[t]
\includegraphics[width=\linewidth]{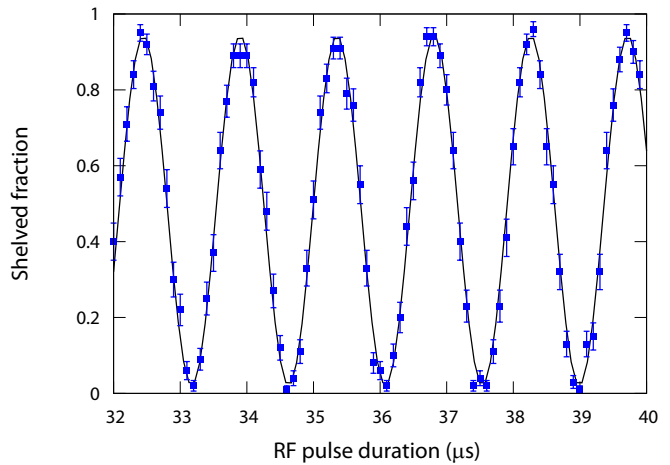}
\caption{Example of Zeeman qubit rotations of the ground state.  After initialization to a single Zeeman sublevel of the ground state, the rf signal is applied for the indicated time followed by $\approx97\%$ efficient transfer of the $m_J=-\frac{1}{2}$ ground state to the $D_{5/2}$ level.  Thus, a shelving efficiency near 1 corresponds to the ion in the $m_J = -\frac{1}{2}$ ground state, while a shelving efficiency near 0 corresponds to the ion in the ${m_J = +\frac{1}{2}}$ ground state.  (Color online.)}
\label{ZeemanFlops}
\end{figure}

\paragraph*{}  A dedicated wire loop in the vicinity of the trap inside the vacuum chamber was used to generate a tunable radio frequency (rf) magnetic field in order to drive magnetic dipole transitions between the Zeeman sublevels for the necessary ion qubit rotations.  An rf synthesizer (based on a direct digital synthesis (DDS9910) chip) generated a stable sinusoidal voltage which was amplified and dropped over a 50 ohm rf resistor for impedance matching, producing an oscillating current in the loop.  This setup allowed for ion qubit rotations with a tunable phase.  An example of ground state Rabi oscillations is shown in Fig. \ref{ZeemanFlops}.  The reduction in contrast is due to imperfect optical pumping ($\approx 98\%$) and shelving efficiency ($\approx 97\%$).

\paragraph*{}  Spontaneously emitted photons on the $6P_{1/2}$ to $6S_{1/2}$ transtion were collected through an $f/2.8$ lens while other wavelengths were extinguished by an interference filter.  Photon collection occurred along a direction perpendicular to the quantization axis.  The two transitions of note in this experiment, $\sigma^-$ and $\pi$, are emitted into the dipole radiation patterns ${|\sigma^-\rangle = \frac{e^{-i\phi}}{\sqrt{2}}(\cos(\theta)|\hat{\theta}\rangle - i|\hat{\phi}\rangle)}$ and ${|\pi\rangle = -\sin(\theta)|\hat{\theta}\rangle}$.  Thus, perpendicular to the quantization axis, i.e. $\theta=\frac{\pi}{2}$, the two polarizations are linear and orthogonal, with their relative intensity differing by a factor of $2$.  A $\lambda/2$ waveplate along the imaging axis allowed for rotations of the photonic qubit.  After passing through a polarizing beamsplitter (PBS), where $\sigma^-$ and $\pi$ correspond to S and P polarizations, respectively, in the unrotated photon basis, the photons were detected at one of two photomultiplier tubes (PMTs) each with quantum\nopagebreak[4] effic\nopagebreak iency~\nopagebreak[4]$\eta\approx0.2$.

\section{Experimental Procedure}

\begin{figure}[b]
\includegraphics[width=\linewidth]{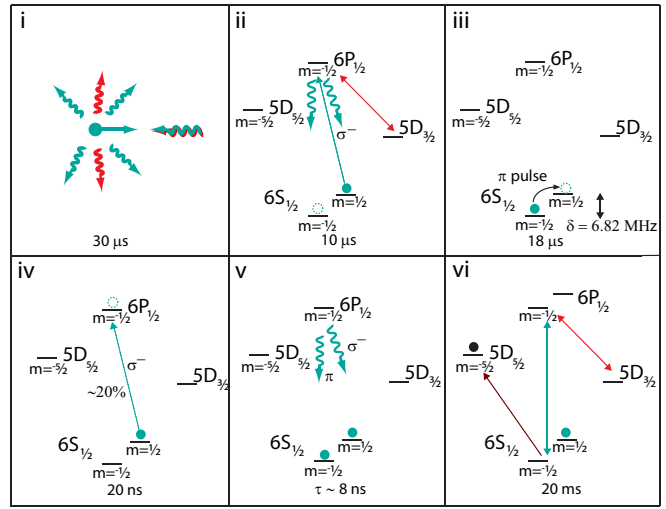}
\caption{Experimental procedure: (i) the ion is Doppler cooled for 30 $\mu$s by 493 nm and 650 nm beams; (ii) state initialization is performed with 10 $\mu$s of optical pumping by a $\sigma^-$-polarized 493 nm beam followed by (iii) an 18 $\mu$s rf $\pi$-pulse resonant with the $6.82$ MHz ground state separation; (iv) the ion is weakly excited with the $\sigma^-$-polarized beam; (v) the excited ion decays emitting a $\sigma^-$-polarized or $\pi$-polarized photon, resulting in the corresponding ground state sublevel, and (vi) ion detection is performed using the 1762 nm shelving beam, followed by turning on the cooling lasers.  (Color online.)}
\label{DutyCycle}
\end{figure}

\paragraph*{}  The experimental cycle, resulting states, and the energy level diagram for \Ba are detailed in Fig. \ref{DutyCycle}.  The experimental cycle ($\approx17 \textrm{ kHz}$ rep rate) consisted of four major steps.  First the ion was Doppler cooled for $30 \textrm{ } \mu$s.  Next the ion was pumped to the $6S_{1/2}$ ($m_J = -\frac{1}{2}$) ground state sublevel with $10 \textrm{ } \mu$s of the circularly polarized optical pumping beam.  State initialization was completed by following the optical pumping with a resonant rf $\pi$-pulse to transfer the ion state to $6S_{1/2}$ ($m_J=+\frac{1}{2}$).  Finally the ion was weakly excited (excitation probability $\approx20\%$) to the $6P_{1/2}$ ($m_J = -\frac{1}{2}$) sublevel (lifetime $\approx 8 \textrm{ ns}$ \cite {Pinnington95}) with a 20 ns exposure to the optical pumping beam.  This was followed by photon- and ion-state detection.  The photons were detected by the PMTs inside a detection window of 20 ns, delayed somewhat from the excitation pulse.  The excitation pulse length, detection window, and its delay were optimized to reduce background counts and double excitations while maintaining a reasonable experimental rate.  Following a single photon detection, the ion state was determined by a shelving pulse followed by Doppler cooling.  ``Dark" state detection efficiency was $\approx 97\%$ limited by the shelving efficiency, and ``bright" state detection efficiency was $> 99\%$.

\paragraph*{}  The resultant entangled state from the spontaneous decay of the ion is ${\frac{1}{\sqrt{2}}|\sigma^-\uparrow\rangle + \frac{1}{\sqrt{2}}|\pi\downarrow\rangle}$ as it decays to either the $6S_{1/2}$ ($m_J = +\frac{1}{2}$) or $6S_{1/2}$ ($m_J = -\frac{1}{2}$) sublevel while emitting either a $\sigma^-$ or $\pi$ photon respectively with equal probability.   However, along the imaging axis, the radiation patterns of the two polarizations differ by a factor of two.  Thus the ideal measured state is ${\frac{1}{\sqrt{3}}|\sigma^-\uparrow\rangle + \sqrt{\frac{2}{3}}|\pi\downarrow\rangle}$ whose overlap with a Bell state is $0.97$.

\paragraph*{}  We first measured the correlation between the ionic and photonic qubits by varying the photonic measurement basis with the $\lambda/2$ waveplate.  Without rotating the ionic qubit, the state probabilities are proportional to $\sin^2(\frac{\theta_{photon}}{2})$ or $\cos^2(\frac{\theta_{photon}}{2})$, thus correlation fringes are produced with varying $\theta_{photon}$.  At each waveplate setting, 100 runs were taken, and the conditional probabilities were plotted in Fig. \ref{Correlations} (i).  At the rotation angle corresponding to maximum correlation, 1000 runs were taken for better statistics in the fidelity calculation.

\paragraph*{}  Then to verify entanglement, the ionic and photonic qubits were both rotated by $\frac{\pi}{2}$ on the Bloch sphere, and the correlation was measured while varying the phase of the ionic qubit rotation, $\phi_I$. The relative phase of the ionic versus photonic qubit rotations is given by ${\phi = \delta\tau + \phi_I}$, where $\delta = 2\pi(6.82 \textrm{ MHz})$ is the ground state Zeeman splitting and $\tau$ is the time delay between emission of the photon and application of the ionic qubit rotation.  If one assumes that $\delta\tau$ is constant, then varying $\phi_I$ will produce correlation fringes since the state probabilities are proportional to $1\pm\cos{\phi}$.  At each $\phi_I$ setting, 500 runs were taken, and the conditional probabilities were plotted in Fig. \ref{Correlations} (ii).  At the maximum correlation, 2000 runs were taken for better statistics in the fidelity calculation.  
Following these measurements, a CHSH measurement was made with a relative phase $\phi = 0$ using those rotations resulting in a maximal violation of Eq. \ref{CHSH}.

\paragraph*{}  The entanglement generation rate depends on the probability of success $P$ of a single experimental run which relies on the excitation probability and the probability of detecting the spontaneously emitted photon.  Given an excitation probability ($P_{exc}\approx0.2$) along with the branching ratio to the ground state ($f=0.75$), the quantum efficiency of the PMTs ($\eta\approx0.2$), fraction of solid angle of light collection $(\frac{\Delta\Omega}{4\pi}\approx0.02$), fraction of photons within PMT detection window ($f_{gate}\approx0.8$), and the transmission losses ($T\approx0.3$) through the imaging optics including an interference filter with $50\%$ transmission, we expect the experimental entanglement generation rate to be $R = P(17 \textrm{ kHz}) = P_{exc}f\eta(\frac{\Delta\Omega}{4\pi})f_{gate}T(17 \textrm{ kHz}) = (0.2)(0.75)(0.2)(0.02)(0.8)(0.3)(17 \textrm{ kHz}) = 2.5 \textrm{ Hz}$ which agrees well with our results.

\section{Results}

\begin{figure}[ht]
\includegraphics[width=\linewidth]{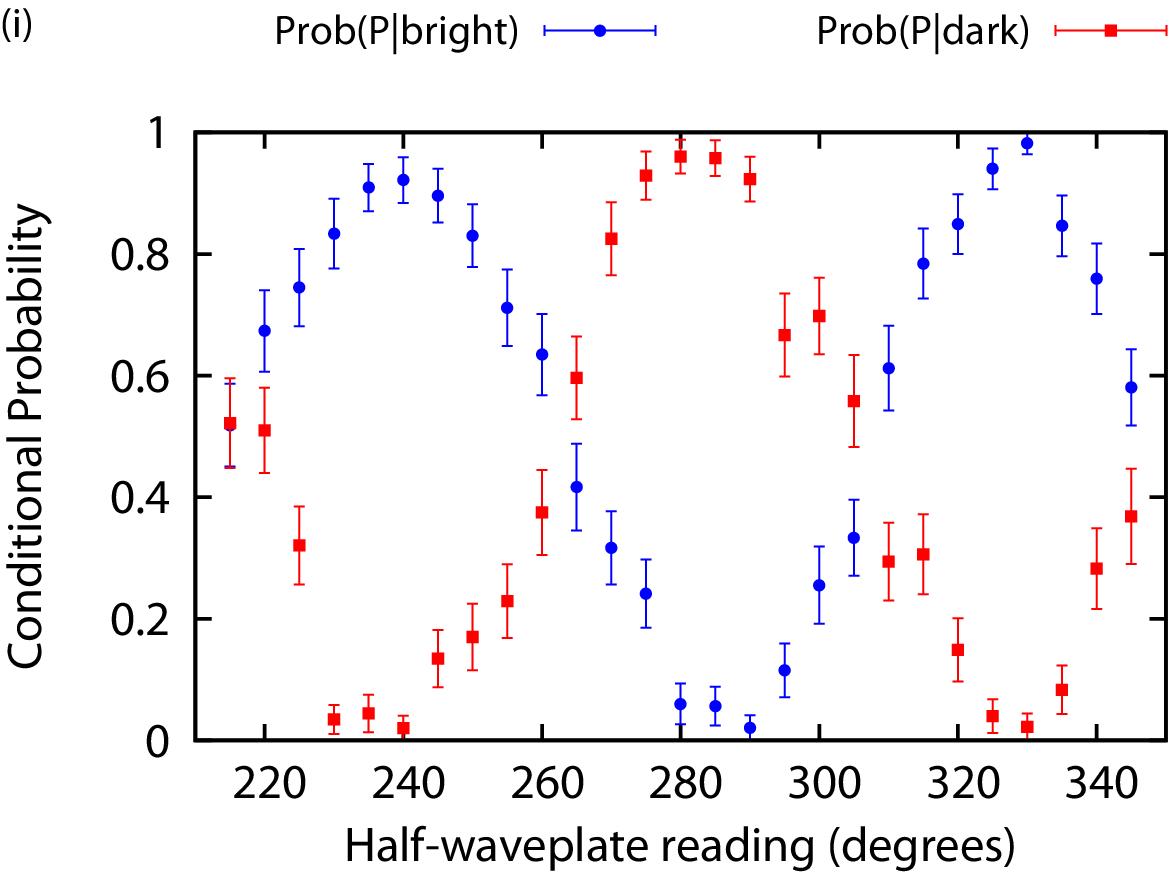}

\vspace{8mm}

\includegraphics[width=\linewidth]{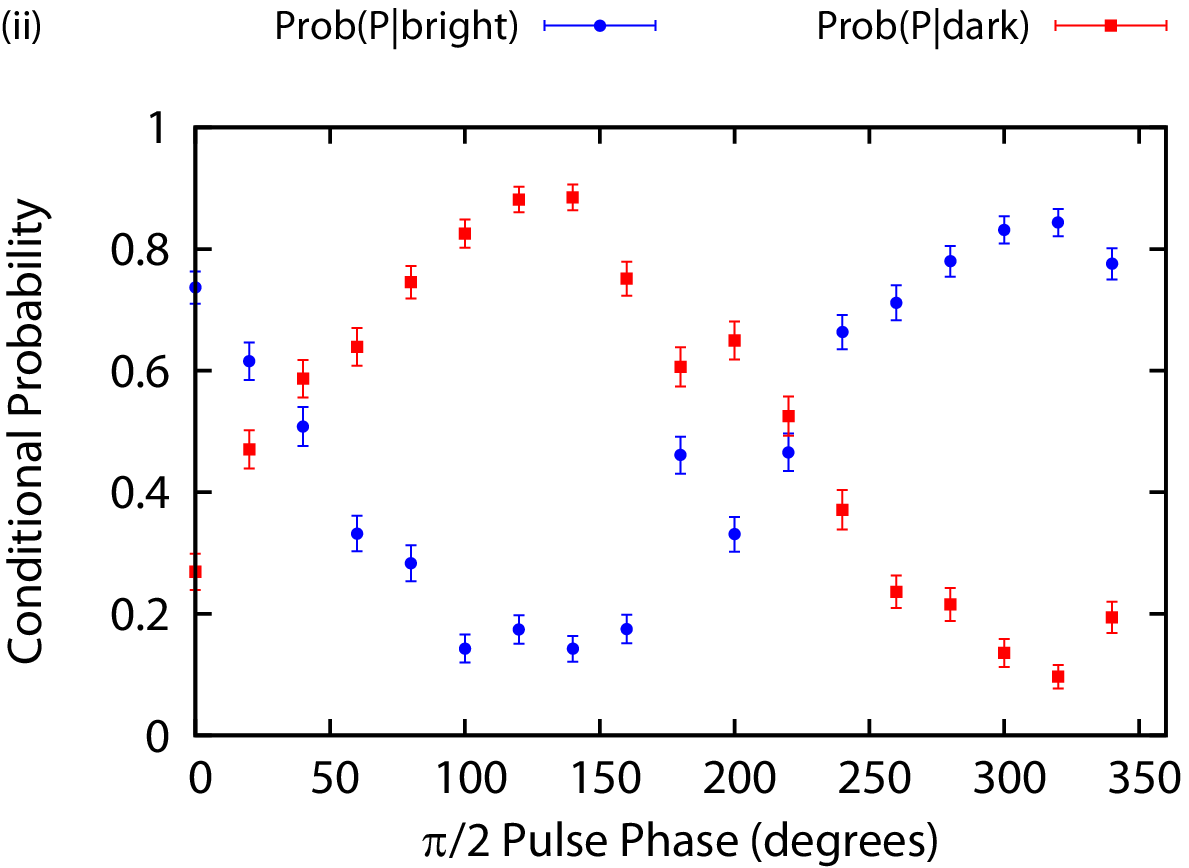}
\caption{Measured conditional probabilities $P(\textrm{P}|\textrm{bright})$ and $P(\textrm{P}|\textrm{dark})$ (i) as a function of the photon qubit measurement basis angle by rotating the half-waveplate while leaving the ion qubit unrotated, and (ii) versus the ion-quibt phase, $\phi_I$, between the rotations for rotation of each qubit by a polar angle of $\frac{\pi}{2}$.  (Color online.)}
\label{Correlations}
\end{figure}

\paragraph*{}  The correlation fringes in the measured conditional probabilities of the photonic and ionic qubit states are plotted in Fig. \ref{Correlations}.  We find a maximal correlation of ${P(\textrm{P}|\textrm{bright}) = 0.05 \pm 0.01}$, ${P(\textrm{P}|\textrm{dark}) = 0.95 \pm 0.01}$, ${P(\textrm{S}|\textrm{bright}) = 0.97 \pm 0.01}$, and ${P(\textrm{S}|\textrm{dark}) = 0.03 \pm 0.01}$ for the unrotated bases and ${P(\textrm{P}|\textrm{bright}) = 0.12 \pm 0.01}$, ${P(\textrm{P}|\textrm{dark}) = 0.88 \pm 0.01}$, ${P(\textrm{S}|\textrm{bright}) = 0.88 \pm 0.01}$, and ${P(\textrm{S}|\textrm{dark}) = 0.12 \pm 0.01}$ for the rotated bases where the errors given are statistical for 1000 and 2000 successful trials, respectively.  The correlations in the rotated basis are weaker largely due to timing jitter in the application of the ion state rotation.  From this we use Eq. \ref{Fidelity2} to calculate the entanglement fidelity to be ${F \geq 0.84\pm0.01}$, which clearly exceeds the limit of ${F > 0.5}$ for entanglement \cite{Sackett00}.

\begin{table}[b]
\caption{\label{tab:Bellsig}
Correlation function measurements and calculated Bell Signals. }
\begin{ruledtabular}
\begin{tabular}{ccc}
\textrm{$\theta_{ion}$}&
\textrm{$\theta_{photon}$}&
\textrm{$P(\theta_{ion}, \theta_{photon})$}\\

\colrule
$0$ &  $\frac{\pi}{4}$ & $0.549$ \\
$0$ & $\frac{3\pi}{4}$ & $-0.607$\\
$\frac{\pi}{2}$ & $\frac{\pi}{4}$ & $0.621$\\
$\frac{\pi}{2}$ & $\frac{3\pi}{4}$ & $0.516$\\
& $S = 2.293 \pm 0.036$ & \\
\hline
$\frac{\pi}{4}$ &  $0$ & $0.654$ \\
$\frac{3\pi}{4}$ & $0$ & $-0.579$\\
$\frac{\pi}{4}$ & $\frac{\pi}{2}$ & $0.542$\\
$\frac{3\pi}{4}$ & $\frac{\pi}{2}$ & $0.528$\\
& $S = 2.303 \pm 0.036$ & \\

\end{tabular}
\end{ruledtabular}
\end{table}

\paragraph*{}  The CHSH measurement was made using the maximally violating qubit rotation angles $\theta_{a,a'} = 0$, $\frac{\pi}{2}$ and $\theta_{b,b'} = \frac{\pi}{4}$, $\frac{3\pi}{4}$.  The four correlations were measured in two ways: first using the $a,a'$ rotations for the ion qubit, and second using the $a,a'$ rotations for the photonic qubit.  $2000$ successful entanglement events make up each correlation measurement.  From these measurements, we find a Bell signal of ${S = 2.293 \pm 0.036}$ for the first case and ${S = 2.303 \pm 0.036}$ for the second case.  The errors are statistical and the Bell signal exceeds $2$ by over eight standard deviations.  A summary of these results is displayed in Table \ref{tab:Bellsig}.  Note that based on the fidelity of the entangled state, we use Eq. \ref{Bellsignal} to calculate an expected Bell signal to be ${S = 2.373\pm0.027}$.

\paragraph*{}  The fidelity and Bell signal are reduced by several factors.  Multiple excitations ($5\%$), imperfect rotation of the ionic qubit due to different arrival times of photons within the 20 ns PMT detection window ($3\%$), mixing of photon polarizations ($0.5\%$), background and dark counts ($3\textrm{--}5\%$), imperfect PBS ($4\%$), and magnetic field fluctuations ($6\%$).

\section{Conclusions}

\paragraph*{}  The ion-photon entanglement demonstrated here allows for the construction of a quantum network over a large distance.  Especially interesting would be a loophole free Bell inequality test using photon mediated entanglement swapping between distant ion-photon entangled pairs.  \Ba  is especially suited for this due to the high fidelity of ion state detection and the relatively low attenuation of 493 nm light over fiber optic cable.  In this case high efficiency of the collection of photons from one entangled ion-photon pair would significantly increase the experimental rate of ion-ion entanglement generation.  Using a parabolic mirror trap, our group has recently attained as much as $40\%$ light collection efficiency from a single trapped ion \cite{Chou13}.  Future work includes improvement of entanglement fidelity by the use of ultrafast pulses in order to approach unit excitation with a single pulse, thus eliminating multiple excitations and background counts.

\begin{acknowledgments}
The authors wish to thank John Wright, Matt Hoffman, Tomasz Sakrejda, Spencer Williams, Richard Graham, Zichao Zhou, and Anupriya Jayakumar for useful conversations.  The authors would also like to acknowledge the preceding work of Matt Dietrich, Nathan Kurz, Paul Pham, and Ryan Bowler that contributed to this experiment.  This research was supported by National Science Foundation  Grant No. 0904004.
\end{acknowledgments}


\begin{thebibliography}{32}%
\makeatletter
\providecommand \@ifxundefined [1]{%
 \@ifx{#1\undefined}
}%
\providecommand \@ifnum [1]{%
 \ifnum #1\expandafter \@firstoftwo
 \else \expandafter \@secondoftwo
 \fi
}%
\providecommand \@ifx [1]{%
 \ifx #1\expandafter \@firstoftwo
 \else \expandafter \@secondoftwo
 \fi
}%
\providecommand \natexlab [1]{#1}%
\providecommand \enquote  [1]{``#1''}%
\providecommand \bibnamefont  [1]{#1}%
\providecommand \bibfnamefont [1]{#1}%
\providecommand \citenamefont [1]{#1}%
\providecommand \href@noop [0]{\@secondoftwo}%
\providecommand \href [0]{\begingroup \@sanitize@url \@href}%
\providecommand \@href[1]{\@@startlink{#1}\@@href}%
\providecommand \@@href[1]{\endgroup#1\@@endlink}%
\providecommand \@sanitize@url [0]{\catcode `\\12\catcode `\$12\catcode
  `\&12\catcode `\#12\catcode `\^12\catcode `\_12\catcode `\%12\relax}%
\providecommand \@@startlink[1]{}%
\providecommand \@@endlink[0]{}%
\providecommand \url  [0]{\begingroup\@sanitize@url \@url }%
\providecommand \@url [1]{\endgroup\@href {#1}{\urlprefix }}%
\providecommand \urlprefix  [0]{URL }%
\providecommand \Eprint [0]{\href }%
\providecommand \doibase [0]{http://dx.doi.org/}%
\providecommand \selectlanguage [0]{\@gobble}%
\providecommand \bibinfo  [0]{\@secondoftwo}%
\providecommand \bibfield  [0]{\@secondoftwo}%
\providecommand \translation [1]{[#1]}%
\providecommand \BibitemOpen [0]{}%
\providecommand \bibitemStop [0]{}%
\providecommand \bibitemNoStop [0]{.\EOS\space}%
\providecommand \EOS [0]{\spacefactor3000\relax}%
\providecommand \BibitemShut  [1]{\csname bibitem#1\endcsname}%
\let\auto@bib@innerbib\@empty
\bibitem [{\citenamefont {De~Greve}\ \emph {et~al.}(2012)\citenamefont
  {De~Greve}, \citenamefont {Yu}, \citenamefont {McMahon}, \citenamefont
  {Pelc}, \citenamefont {Natarajan}, \citenamefont {Kim}, \citenamefont {Abe},
  \citenamefont {Maier}, \citenamefont {Schneider}, \citenamefont {Kamp},
  \citenamefont {Hofling}, \citenamefont {Hadfield}, \citenamefont {Forchel},
  \citenamefont {Fejer},\ and\ \citenamefont {Yamamoto}}]{DeGreve13}%
  \BibitemOpen
  \bibfield  {author} {\bibinfo {author} {\bibfnamefont {K.}~\bibnamefont
  {De~Greve}}, \bibinfo {author} {\bibfnamefont {L.}~\bibnamefont {Yu}},
  \bibinfo {author} {\bibfnamefont {P.~L.}\ \bibnamefont {McMahon}}, \bibinfo
  {author} {\bibfnamefont {J.~S.}\ \bibnamefont {Pelc}}, \bibinfo {author}
  {\bibfnamefont {C.~M.}\ \bibnamefont {Natarajan}}, \bibinfo {author}
  {\bibfnamefont {N.~Y.}\ \bibnamefont {Kim}}, \bibinfo {author} {\bibfnamefont
  {E.}~\bibnamefont {Abe}}, \bibinfo {author} {\bibfnamefont {S.}~\bibnamefont
  {Maier}}, \bibinfo {author} {\bibfnamefont {C.}~\bibnamefont {Schneider}},
  \bibinfo {author} {\bibfnamefont {M.}~\bibnamefont {Kamp}}, \bibinfo {author}
  {\bibfnamefont {S.}~\bibnamefont {Hofling}}, \bibinfo {author} {\bibfnamefont
  {R.~H.}\ \bibnamefont {Hadfield}}, \bibinfo {author} {\bibfnamefont
  {A.}~\bibnamefont {Forchel}}, \bibinfo {author} {\bibfnamefont {M.~M.}\
  \bibnamefont {Fejer}}, \ and\ \bibinfo {author} {\bibfnamefont
  {Y.}~\bibnamefont {Yamamoto}},\ }\href {\doibase 10.1038/nature11577}
  {\bibfield  {journal} {\bibinfo  {journal} {Nature}\ }\textbf {\bibinfo
  {volume} {491}},\ \bibinfo {pages} {421} (\bibinfo {year}
  {2012})}\BibitemShut {NoStop}%
\bibitem [{\citenamefont {Gao}\ \emph {et~al.}(2012)\citenamefont {Gao},
  \citenamefont {Fallahi}, \citenamefont {Togan}, \citenamefont
  {Miguel-Sanchez},\ and\ \citenamefont {Imamoglu}}]{Gao12}%
  \BibitemOpen
  \bibfield  {author} {\bibinfo {author} {\bibfnamefont {W.~B.}\ \bibnamefont
  {Gao}}, \bibinfo {author} {\bibfnamefont {P.}~\bibnamefont {Fallahi}},
  \bibinfo {author} {\bibfnamefont {E.}~\bibnamefont {Togan}}, \bibinfo
  {author} {\bibfnamefont {J.}~\bibnamefont {Miguel-Sanchez}}, \ and\ \bibinfo
  {author} {\bibfnamefont {A.}~\bibnamefont {Imamoglu}},\ }\href {\doibase
  10.1038/nature11573} {\bibfield  {journal} {\bibinfo  {journal} {Nature}\
  }\textbf {\bibinfo {volume} {491}},\ \bibinfo {pages} {426} (\bibinfo {year}
  {2012})}\BibitemShut {NoStop}%
\bibitem [{\citenamefont {Juska}\ \emph {et~al.}(2013)\citenamefont {Juska},
  \citenamefont {Dimastrodonato}, \citenamefont {Mereni}, \citenamefont
  {Gocalinska},\ and\ \citenamefont {Pelucchi}}]{Juska13}%
  \BibitemOpen
  \bibfield  {author} {\bibinfo {author} {\bibfnamefont {G.}~\bibnamefont
  {Juska}}, \bibinfo {author} {\bibfnamefont {V.}~\bibnamefont
  {Dimastrodonato}}, \bibinfo {author} {\bibfnamefont {L.~O.}\ \bibnamefont
  {Mereni}}, \bibinfo {author} {\bibfnamefont {A.}~\bibnamefont {Gocalinska}},
  \ and\ \bibinfo {author} {\bibfnamefont {E.}~\bibnamefont {Pelucchi}},\
  }\href {\doibase 10.1038/nphoton.2013.128} {\bibfield  {journal} {\bibinfo
  {journal} {Nat Photon}\ }\textbf {\bibinfo {volume} {7}},\ \bibinfo {pages}
  {527} (\bibinfo {year} {2013})}\BibitemShut {NoStop}%
\bibitem [{\citenamefont {Togan}\ \emph {et~al.}(2010)\citenamefont {Togan},
  \citenamefont {Chu}, \citenamefont {Trifonov}, \citenamefont {Jiang},
  \citenamefont {Maze}, \citenamefont {Childress}, \citenamefont {Dutt},
  \citenamefont {Sorensen}, \citenamefont {Hemmer}, \citenamefont {Zibrov},\
  and\ \citenamefont {Lukin}}]{Togan10}%
  \BibitemOpen
  \bibfield  {author} {\bibinfo {author} {\bibfnamefont {E.}~\bibnamefont
  {Togan}}, \bibinfo {author} {\bibfnamefont {Y.}~\bibnamefont {Chu}}, \bibinfo
  {author} {\bibfnamefont {A.~S.}\ \bibnamefont {Trifonov}}, \bibinfo {author}
  {\bibfnamefont {L.}~\bibnamefont {Jiang}}, \bibinfo {author} {\bibfnamefont
  {J.}~\bibnamefont {Maze}}, \bibinfo {author} {\bibfnamefont {L.}~\bibnamefont
  {Childress}}, \bibinfo {author} {\bibfnamefont {M.~V.~G.}\ \bibnamefont
  {Dutt}}, \bibinfo {author} {\bibfnamefont {A.~S.}\ \bibnamefont {Sorensen}},
  \bibinfo {author} {\bibfnamefont {P.~R.}\ \bibnamefont {Hemmer}}, \bibinfo
  {author} {\bibfnamefont {A.~S.}\ \bibnamefont {Zibrov}}, \ and\ \bibinfo
  {author} {\bibfnamefont {M.~D.}\ \bibnamefont {Lukin}},\ }\href {\doibase
  10.1038/nature09256} {\bibfield  {journal} {\bibinfo  {journal} {Nature}\
  }\textbf {\bibinfo {volume} {466}},\ \bibinfo {pages} {730} (\bibinfo {year}
  {2010})}\BibitemShut {NoStop}%
\bibitem [{\citenamefont {Neumann}\ \emph {et~al.}(2008)\citenamefont
  {Neumann}, \citenamefont {Mizuochi}, \citenamefont {Rempp}, \citenamefont
  {Hemmer}, \citenamefont {Watanabe}, \citenamefont {Yamasaki}, \citenamefont
  {Jacques}, \citenamefont {Gaebel}, \citenamefont {Jelezko},\ and\
  \citenamefont {Wrachtrup}}]{Neumann08}%
  \BibitemOpen
  \bibfield  {author} {\bibinfo {author} {\bibfnamefont {P.}~\bibnamefont
  {Neumann}}, \bibinfo {author} {\bibfnamefont {N.}~\bibnamefont {Mizuochi}},
  \bibinfo {author} {\bibfnamefont {F.}~\bibnamefont {Rempp}}, \bibinfo
  {author} {\bibfnamefont {P.}~\bibnamefont {Hemmer}}, \bibinfo {author}
  {\bibfnamefont {H.}~\bibnamefont {Watanabe}}, \bibinfo {author}
  {\bibfnamefont {S.}~\bibnamefont {Yamasaki}}, \bibinfo {author}
  {\bibfnamefont {V.}~\bibnamefont {Jacques}}, \bibinfo {author} {\bibfnamefont
  {T.}~\bibnamefont {Gaebel}}, \bibinfo {author} {\bibfnamefont
  {F.}~\bibnamefont {Jelezko}}, \ and\ \bibinfo {author} {\bibfnamefont
  {J.}~\bibnamefont {Wrachtrup}},\ }\href {\doibase 10.1126/science.1157233}
  {\bibfield  {journal} {\bibinfo  {journal} {Science}\ }\textbf {\bibinfo
  {volume} {320}},\ \bibinfo {pages} {1326} (\bibinfo {year}
  {2008})}\BibitemShut {NoStop}%
\bibitem [{\citenamefont {Bernien}\ \emph {et~al.}(2012)\citenamefont
  {Bernien}, \citenamefont {Childress}, \citenamefont {Robledo}, \citenamefont
  {Markham}, \citenamefont {Twitchen},\ and\ \citenamefont
  {Hanson}}]{Bernien12}%
  \BibitemOpen
  \bibfield  {author} {\bibinfo {author} {\bibfnamefont {H.}~\bibnamefont
  {Bernien}}, \bibinfo {author} {\bibfnamefont {L.}~\bibnamefont {Childress}},
  \bibinfo {author} {\bibfnamefont {L.}~\bibnamefont {Robledo}}, \bibinfo
  {author} {\bibfnamefont {M.}~\bibnamefont {Markham}}, \bibinfo {author}
  {\bibfnamefont {D.}~\bibnamefont {Twitchen}}, \ and\ \bibinfo {author}
  {\bibfnamefont {R.}~\bibnamefont {Hanson}},\ }\href {\doibase
  10.1103/PhysRevLett.108.043604} {\bibfield  {journal} {\bibinfo  {journal}
  {Phys. Rev. Lett.}\ }\textbf {\bibinfo {volume} {108}},\ \bibinfo {pages}
  {043604} (\bibinfo {year} {2012})}\BibitemShut {NoStop}%
\bibitem [{\citenamefont {Wilk}\ \emph {et~al.}(2010)\citenamefont {Wilk},
  \citenamefont {Ga\"etan}, \citenamefont {Evellin}, \citenamefont {Wolters},
  \citenamefont {Miroshnychenko}, \citenamefont {Grangier},\ and\ \citenamefont
  {Browaeys}}]{Wilk10}%
  \BibitemOpen
  \bibfield  {author} {\bibinfo {author} {\bibfnamefont {T.}~\bibnamefont
  {Wilk}}, \bibinfo {author} {\bibfnamefont {A.}~\bibnamefont {Ga\"etan}},
  \bibinfo {author} {\bibfnamefont {C.}~\bibnamefont {Evellin}}, \bibinfo
  {author} {\bibfnamefont {J.}~\bibnamefont {Wolters}}, \bibinfo {author}
  {\bibfnamefont {Y.}~\bibnamefont {Miroshnychenko}}, \bibinfo {author}
  {\bibfnamefont {P.}~\bibnamefont {Grangier}}, \ and\ \bibinfo {author}
  {\bibfnamefont {A.}~\bibnamefont {Browaeys}},\ }\href {\doibase
  10.1103/PhysRevLett.104.010502} {\bibfield  {journal} {\bibinfo  {journal}
  {Phys. Rev. Lett.}\ }\textbf {\bibinfo {volume} {104}},\ \bibinfo {pages}
  {010502} (\bibinfo {year} {2010})}\BibitemShut {NoStop}%
\bibitem [{\citenamefont {Rosenfeld}\ \emph {et~al.}(2008)\citenamefont
  {Rosenfeld}, \citenamefont {Hocke}, \citenamefont {Henkel}, \citenamefont
  {Krug}, \citenamefont {Volz}, \citenamefont {Weber},\ and\ \citenamefont
  {Weinfurter}}]{Rosenfeld08}%
  \BibitemOpen
  \bibfield  {author} {\bibinfo {author} {\bibfnamefont {W.}~\bibnamefont
  {Rosenfeld}}, \bibinfo {author} {\bibfnamefont {F.}~\bibnamefont {Hocke}},
  \bibinfo {author} {\bibfnamefont {F.}~\bibnamefont {Henkel}}, \bibinfo
  {author} {\bibfnamefont {M.}~\bibnamefont {Krug}}, \bibinfo {author}
  {\bibfnamefont {J.}~\bibnamefont {Volz}}, \bibinfo {author} {\bibfnamefont
  {M.}~\bibnamefont {Weber}}, \ and\ \bibinfo {author} {\bibfnamefont
  {H.}~\bibnamefont {Weinfurter}},\ }\href {\doibase
  10.1103/PhysRevLett.101.260403} {\bibfield  {journal} {\bibinfo  {journal}
  {Phys. Rev. Lett.}\ }\textbf {\bibinfo {volume} {101}},\ \bibinfo {pages}
  {260403} (\bibinfo {year} {2008})}\BibitemShut {NoStop}%
\bibitem [{\citenamefont {Li}\ \emph {et~al.}(2013)\citenamefont {Li},
  \citenamefont {Dudin},\ and\ \citenamefont {Kuzmich}}]{Li13}%
  \BibitemOpen
  \bibfield  {author} {\bibinfo {author} {\bibfnamefont {L.}~\bibnamefont
  {Li}}, \bibinfo {author} {\bibfnamefont {Y.~O.}\ \bibnamefont {Dudin}}, \
  and\ \bibinfo {author} {\bibfnamefont {A.}~\bibnamefont {Kuzmich}},\ }\href
  {\doibase 10.1038/nature12227} {\bibfield  {journal} {\bibinfo  {journal}
  {Nature}\ }\textbf {\bibinfo {volume} {498}},\ \bibinfo {pages} {466}
  (\bibinfo {year} {2013})}\BibitemShut {NoStop}%
\bibitem [{\citenamefont {Chou}\ \emph {et~al.}(2005)\citenamefont {Chou},
  \citenamefont {de~Riedmatten}, \citenamefont {Felinto}, \citenamefont
  {Polyakov}, \citenamefont {van Enk},\ and\ \citenamefont {Kimble}}]{Chou05}%
  \BibitemOpen
  \bibfield  {author} {\bibinfo {author} {\bibfnamefont {C.~W.}\ \bibnamefont
  {Chou}}, \bibinfo {author} {\bibfnamefont {H.}~\bibnamefont {de~Riedmatten}},
  \bibinfo {author} {\bibfnamefont {D.}~\bibnamefont {Felinto}}, \bibinfo
  {author} {\bibfnamefont {S.~V.}\ \bibnamefont {Polyakov}}, \bibinfo {author}
  {\bibfnamefont {J.}~\bibnamefont {van Enk}}, \ and\ \bibinfo {author}
  {\bibfnamefont {H.~J.}\ \bibnamefont {Kimble}},\ }\href {\doibase
  10.1038/nature04353} {\bibfield  {journal} {\bibinfo  {journal} {Nature}\
  }\textbf {\bibinfo {volume} {438}},\ \bibinfo {pages} {828} (\bibinfo {year}
  {2005})}\BibitemShut {NoStop}%
\bibitem [{\citenamefont {Eichler}\ \emph {et~al.}(2012)\citenamefont
  {Eichler}, \citenamefont {Lang}, \citenamefont {Fink}, \citenamefont
  {Govenius}, \citenamefont {Filipp},\ and\ \citenamefont
  {Wallraff}}]{Eichler12}%
  \BibitemOpen
  \bibfield  {author} {\bibinfo {author} {\bibfnamefont {C.}~\bibnamefont
  {Eichler}}, \bibinfo {author} {\bibfnamefont {C.}~\bibnamefont {Lang}},
  \bibinfo {author} {\bibfnamefont {J.~M.}\ \bibnamefont {Fink}}, \bibinfo
  {author} {\bibfnamefont {J.}~\bibnamefont {Govenius}}, \bibinfo {author}
  {\bibfnamefont {S.}~\bibnamefont {Filipp}}, \ and\ \bibinfo {author}
  {\bibfnamefont {A.}~\bibnamefont {Wallraff}},\ }\href {\doibase
  10.1103/PhysRevLett.109.240501} {\bibfield  {journal} {\bibinfo  {journal}
  {Phys. Rev. Lett.}\ }\textbf {\bibinfo {volume} {109}},\ \bibinfo {pages}
  {240501} (\bibinfo {year} {2012})}\BibitemShut {NoStop}%
\bibitem [{\citenamefont {Steffen}\ \emph {et~al.}(2006)\citenamefont
  {Steffen}, \citenamefont {Ansmann}, \citenamefont {Bialczak}, \citenamefont
  {Katz}, \citenamefont {Lucero}, \citenamefont {McDermott}, \citenamefont
  {Neeley}, \citenamefont {Weig}, \citenamefont {Cleland},\ and\ \citenamefont
  {Martinis}}]{Steffen06}%
  \BibitemOpen
  \bibfield  {author} {\bibinfo {author} {\bibfnamefont {M.}~\bibnamefont
  {Steffen}}, \bibinfo {author} {\bibfnamefont {M.}~\bibnamefont {Ansmann}},
  \bibinfo {author} {\bibfnamefont {R.~C.}\ \bibnamefont {Bialczak}}, \bibinfo
  {author} {\bibfnamefont {N.}~\bibnamefont {Katz}}, \bibinfo {author}
  {\bibfnamefont {E.}~\bibnamefont {Lucero}}, \bibinfo {author} {\bibfnamefont
  {R.}~\bibnamefont {McDermott}}, \bibinfo {author} {\bibfnamefont
  {M.}~\bibnamefont {Neeley}}, \bibinfo {author} {\bibfnamefont {E.~M.}\
  \bibnamefont {Weig}}, \bibinfo {author} {\bibfnamefont {A.~N.}\ \bibnamefont
  {Cleland}}, \ and\ \bibinfo {author} {\bibfnamefont {J.~M.}\ \bibnamefont
  {Martinis}},\ }\href {\doibase 10.1126/science.1130886} {\bibfield  {journal}
  {\bibinfo  {journal} {Science}\ }\textbf {\bibinfo {volume} {313}},\ \bibinfo
  {pages} {1423} (\bibinfo {year} {2006})}\BibitemShut {NoStop}%
\bibitem [{\citenamefont {Berkley}\ \emph {et~al.}(2003)\citenamefont
  {Berkley}, \citenamefont {Xu}, \citenamefont {Ramos}, \citenamefont {Gubrud},
  \citenamefont {Strauch}, \citenamefont {Johnson}, \citenamefont {Anderson},
  \citenamefont {Dragt}, \citenamefont {Lobb},\ and\ \citenamefont
  {Wellstood}}]{Berkley03}%
  \BibitemOpen
  \bibfield  {author} {\bibinfo {author} {\bibfnamefont {A.~J.}\ \bibnamefont
  {Berkley}}, \bibinfo {author} {\bibfnamefont {H.}~\bibnamefont {Xu}},
  \bibinfo {author} {\bibfnamefont {R.~C.}\ \bibnamefont {Ramos}}, \bibinfo
  {author} {\bibfnamefont {M.~A.}\ \bibnamefont {Gubrud}}, \bibinfo {author}
  {\bibfnamefont {F.~W.}\ \bibnamefont {Strauch}}, \bibinfo {author}
  {\bibfnamefont {P.~R.}\ \bibnamefont {Johnson}}, \bibinfo {author}
  {\bibfnamefont {J.~R.}\ \bibnamefont {Anderson}}, \bibinfo {author}
  {\bibfnamefont {A.~J.}\ \bibnamefont {Dragt}}, \bibinfo {author}
  {\bibfnamefont {C.~J.}\ \bibnamefont {Lobb}}, \ and\ \bibinfo {author}
  {\bibfnamefont {F.~C.}\ \bibnamefont {Wellstood}},\ }\href {\doibase
  10.1126/science.1084528} {\bibfield  {journal} {\bibinfo  {journal}
  {Science}\ }\textbf {\bibinfo {volume} {300}},\ \bibinfo {pages} {1548}
  (\bibinfo {year} {2003})}\BibitemShut {NoStop}%
\bibitem [{\citenamefont {Moehring}\ \emph {et~al.}(2007)\citenamefont
  {Moehring}, \citenamefont {Maunz}, \citenamefont {Olmschenk}, \citenamefont
  {Younge}, \citenamefont {Matsukevich}, \citenamefont {Duan},\ and\
  \citenamefont {Monroe}}]{Moehring07}%
  \BibitemOpen
  \bibfield  {author} {\bibinfo {author} {\bibfnamefont {D.~L.}\ \bibnamefont
  {Moehring}}, \bibinfo {author} {\bibfnamefont {P.}~\bibnamefont {Maunz}},
  \bibinfo {author} {\bibfnamefont {S.}~\bibnamefont {Olmschenk}}, \bibinfo
  {author} {\bibfnamefont {K.~C.}\ \bibnamefont {Younge}}, \bibinfo {author}
  {\bibfnamefont {D.~N.}\ \bibnamefont {Matsukevich}}, \bibinfo {author}
  {\bibfnamefont {L.-M.}\ \bibnamefont {Duan}}, \ and\ \bibinfo {author}
  {\bibfnamefont {C.}~\bibnamefont {Monroe}},\ }\href {\doibase
  10.1038/nature06118} {\bibfield  {journal} {\bibinfo  {journal} {Nature}\
  }\textbf {\bibinfo {volume} {449}},\ \bibinfo {pages} {68} (\bibinfo {year}
  {2007})}\BibitemShut {NoStop}%
\bibitem [{\citenamefont {Blinov}\ \emph {et~al.}(2004)\citenamefont {Blinov},
  \citenamefont {Moehring}, \citenamefont {Duan},\ and\ \citenamefont
  {Monroe}}]{Blinov04}%
  \BibitemOpen
  \bibfield  {author} {\bibinfo {author} {\bibfnamefont {B.~B.}\ \bibnamefont
  {Blinov}}, \bibinfo {author} {\bibfnamefont {D.~L.}\ \bibnamefont
  {Moehring}}, \bibinfo {author} {\bibfnamefont {L.-M.}\ \bibnamefont {Duan}},
  \ and\ \bibinfo {author} {\bibfnamefont {C.}~\bibnamefont {Monroe}},\ }\href
  {\doibase 10.1038/nature02377} {\bibfield  {journal} {\bibinfo  {journal}
  {Nature}\ }\textbf {\bibinfo {volume} {428}},\ \bibinfo {pages} {153}
  (\bibinfo {year} {2004})}\BibitemShut {NoStop}%
\bibitem [{\citenamefont {Monroe}\ \emph {et~al.}(2012)\citenamefont {Monroe},
  \citenamefont {Raussendorf}, \citenamefont {Ruthven}, \citenamefont {Brown},
  \citenamefont {Maunz}, \citenamefont {Duan},\ and\ \citenamefont
  {Kim}}]{Monroe12}%
  \BibitemOpen
  \bibfield  {author} {\bibinfo {author} {\bibfnamefont {C.}~\bibnamefont
  {Monroe}}, \bibinfo {author} {\bibfnamefont {R.}~\bibnamefont {Raussendorf}},
  \bibinfo {author} {\bibfnamefont {A.}~\bibnamefont {Ruthven}}, \bibinfo
  {author} {\bibfnamefont {K.~R.}\ \bibnamefont {Brown}}, \bibinfo {author}
  {\bibfnamefont {P.}~\bibnamefont {Maunz}}, \bibinfo {author} {\bibfnamefont
  {L.-M.}\ \bibnamefont {Duan}}, \ and\ \bibinfo {author} {\bibfnamefont
  {J.}~\bibnamefont {Kim}},\ }\href {http://arxiv.org/abs/1208.0391} {\bibfield
   {journal} {\bibinfo  {journal} {pre-print}\ } (\bibinfo {year}
  {2012})}\BibitemShut {NoStop}%
\bibitem [{\citenamefont {Einstein}\ \emph {et~al.}(1935)\citenamefont
  {Einstein}, \citenamefont {Podolsky},\ and\ \citenamefont
  {Rosen}}]{Einstein35}%
  \BibitemOpen
  \bibfield  {author} {\bibinfo {author} {\bibfnamefont {A.}~\bibnamefont
  {Einstein}}, \bibinfo {author} {\bibfnamefont {B.}~\bibnamefont {Podolsky}},
  \ and\ \bibinfo {author} {\bibfnamefont {N.}~\bibnamefont {Rosen}},\ }\href
  {\doibase 10.1103/PhysRev.47.777} {\bibfield  {journal} {\bibinfo  {journal}
  {Phys. Rev.}\ }\textbf {\bibinfo {volume} {47}},\ \bibinfo {pages} {777}
  (\bibinfo {year} {1935})}\BibitemShut {NoStop}%
\bibitem [{\citenamefont {Bell}(1964)}]{Bell64}%
  \BibitemOpen
  \bibfield  {author} {\bibinfo {author} {\bibfnamefont {J.~S.}\ \bibnamefont
  {Bell}},\ }\href@noop {} {\bibfield  {journal} {\bibinfo  {journal}
  {Physics}\ }\textbf {\bibinfo {volume} {1}},\ \bibinfo {pages} {195}
  (\bibinfo {year} {1964})}\BibitemShut {NoStop}%
\bibitem [{\citenamefont {Giustina}\ \emph {et~al.}(2013)\citenamefont
  {Giustina}, \citenamefont {Mech}, \citenamefont {Ramelow}, \citenamefont
  {Wittmann}, \citenamefont {Kofler}, \citenamefont {Beyer}, \citenamefont
  {Lita}, \citenamefont {Calkins}, \citenamefont {Gerrits}, \citenamefont
  {Nam}, \citenamefont {Ursin},\ and\ \citenamefont {Zeilinger}}]{Giustina13}%
  \BibitemOpen
  \bibfield  {author} {\bibinfo {author} {\bibfnamefont {M.}~\bibnamefont
  {Giustina}}, \bibinfo {author} {\bibfnamefont {A.}~\bibnamefont {Mech}},
  \bibinfo {author} {\bibfnamefont {S.}~\bibnamefont {Ramelow}}, \bibinfo
  {author} {\bibfnamefont {B.}~\bibnamefont {Wittmann}}, \bibinfo {author}
  {\bibfnamefont {J.}~\bibnamefont {Kofler}}, \bibinfo {author} {\bibfnamefont
  {J.}~\bibnamefont {Beyer}}, \bibinfo {author} {\bibfnamefont
  {A.}~\bibnamefont {Lita}}, \bibinfo {author} {\bibfnamefont {B.}~\bibnamefont
  {Calkins}}, \bibinfo {author} {\bibfnamefont {T.}~\bibnamefont {Gerrits}},
  \bibinfo {author} {\bibfnamefont {S.~W.}\ \bibnamefont {Nam}}, \bibinfo
  {author} {\bibfnamefont {R.}~\bibnamefont {Ursin}}, \ and\ \bibinfo {author}
  {\bibfnamefont {A.}~\bibnamefont {Zeilinger}},\ }\href {\doibase
  10.1038/nature12012} {\bibfield  {journal} {\bibinfo  {journal} {Nature}\
  }\textbf {\bibinfo {volume} {497}},\ \bibinfo {pages} {227} (\bibinfo {year}
  {2013})}\BibitemShut {NoStop}%
\bibitem [{\citenamefont {Matsukevich}\ \emph {et~al.}(2008)\citenamefont
  {Matsukevich}, \citenamefont {Maunz}, \citenamefont {Moehring}, \citenamefont
  {Olmschenk},\ and\ \citenamefont {Monroe}}]{Matsukevich08}%
  \BibitemOpen
  \bibfield  {author} {\bibinfo {author} {\bibfnamefont {D.~N.}\ \bibnamefont
  {Matsukevich}}, \bibinfo {author} {\bibfnamefont {P.}~\bibnamefont {Maunz}},
  \bibinfo {author} {\bibfnamefont {D.~L.}\ \bibnamefont {Moehring}}, \bibinfo
  {author} {\bibfnamefont {S.}~\bibnamefont {Olmschenk}}, \ and\ \bibinfo
  {author} {\bibfnamefont {C.}~\bibnamefont {Monroe}},\ }\href {\doibase
  10.1103/PhysRevLett.100.150404} {\bibfield  {journal} {\bibinfo  {journal}
  {Phys. Rev. Lett.}\ }\textbf {\bibinfo {volume} {100}},\ \bibinfo {pages}
  {150404} (\bibinfo {year} {2008})}\BibitemShut {NoStop}%
\bibitem [{\citenamefont {Waldherr}\ \emph {et~al.}(2011)\citenamefont
  {Waldherr}, \citenamefont {Neumann}, \citenamefont {Huelga}, \citenamefont
  {Jelezko},\ and\ \citenamefont {Wrachtrup}}]{Waldherr11}%
  \BibitemOpen
  \bibfield  {author} {\bibinfo {author} {\bibfnamefont {G.}~\bibnamefont
  {Waldherr}}, \bibinfo {author} {\bibfnamefont {P.}~\bibnamefont {Neumann}},
  \bibinfo {author} {\bibfnamefont {S.~F.}\ \bibnamefont {Huelga}}, \bibinfo
  {author} {\bibfnamefont {F.}~\bibnamefont {Jelezko}}, \ and\ \bibinfo
  {author} {\bibfnamefont {J.}~\bibnamefont {Wrachtrup}},\ }\href {\doibase
  10.1103/PhysRevLett.107.090401} {\bibfield  {journal} {\bibinfo  {journal}
  {Phys. Rev. Lett.}\ }\textbf {\bibinfo {volume} {107}},\ \bibinfo {pages}
  {090401} (\bibinfo {year} {2011})}\BibitemShut {NoStop}%
\bibitem [{\citenamefont {Ansmann}\ \emph {et~al.}(2009)\citenamefont
  {Ansmann}, \citenamefont {Wang}, \citenamefont {Bialczak}, \citenamefont
  {Hofheinz}, \citenamefont {Lucero}, \citenamefont {Neeley}, \citenamefont
  {O'Connell}, \citenamefont {Sank}, \citenamefont {Weides}, \citenamefont
  {Wenner}, \citenamefont {Cleland},\ and\ \citenamefont
  {Martinis}}]{Ansmann09}%
  \BibitemOpen
  \bibfield  {author} {\bibinfo {author} {\bibfnamefont {M.}~\bibnamefont
  {Ansmann}}, \bibinfo {author} {\bibfnamefont {H.}~\bibnamefont {Wang}},
  \bibinfo {author} {\bibfnamefont {R.~C.}\ \bibnamefont {Bialczak}}, \bibinfo
  {author} {\bibfnamefont {M.}~\bibnamefont {Hofheinz}}, \bibinfo {author}
  {\bibfnamefont {E.}~\bibnamefont {Lucero}}, \bibinfo {author} {\bibfnamefont
  {M.}~\bibnamefont {Neeley}}, \bibinfo {author} {\bibfnamefont {A.~D.}\
  \bibnamefont {O'Connell}}, \bibinfo {author} {\bibfnamefont {D.}~\bibnamefont
  {Sank}}, \bibinfo {author} {\bibfnamefont {M.}~\bibnamefont {Weides}},
  \bibinfo {author} {\bibfnamefont {J.}~\bibnamefont {Wenner}}, \bibinfo
  {author} {\bibfnamefont {A.~N.}\ \bibnamefont {Cleland}}, \ and\ \bibinfo
  {author} {\bibfnamefont {J.~M.}\ \bibnamefont {Martinis}},\ }\href {\doibase
  10.1038/nature08363} {\bibfield  {journal} {\bibinfo  {journal} {Nature}\
  }\textbf {\bibinfo {volume} {461}},\ \bibinfo {pages} {504} (\bibinfo {year}
  {2009})}\BibitemShut {NoStop}%
\bibitem [{\citenamefont {Hasegawa}\ \emph {et~al.}(2003)\citenamefont
  {Hasegawa}, \citenamefont {Loidl}, \citenamefont {Badurek}, \citenamefont
  {Baron},\ and\ \citenamefont {Rauch}}]{Hasegawa03}%
  \BibitemOpen
  \bibfield  {author} {\bibinfo {author} {\bibfnamefont {Y.}~\bibnamefont
  {Hasegawa}}, \bibinfo {author} {\bibfnamefont {R.}~\bibnamefont {Loidl}},
  \bibinfo {author} {\bibfnamefont {G.}~\bibnamefont {Badurek}}, \bibinfo
  {author} {\bibfnamefont {M.}~\bibnamefont {Baron}}, \ and\ \bibinfo {author}
  {\bibfnamefont {H.}~\bibnamefont {Rauch}},\ }\href {\doibase
  10.1038/nature01881} {\bibfield  {journal} {\bibinfo  {journal} {Nature}\
  }\textbf {\bibinfo {volume} {425}},\ \bibinfo {pages} {45} (\bibinfo {year}
  {2003})}\BibitemShut {NoStop}%
\bibitem [{\citenamefont {Walther}\ \emph {et~al.}(2005)\citenamefont
  {Walther}, \citenamefont {Aspelmeyer}, \citenamefont {Resch},\ and\
  \citenamefont {Zeilinger}}]{Walther05}%
  \BibitemOpen
  \bibfield  {author} {\bibinfo {author} {\bibfnamefont {P.}~\bibnamefont
  {Walther}}, \bibinfo {author} {\bibfnamefont {M.}~\bibnamefont {Aspelmeyer}},
  \bibinfo {author} {\bibfnamefont {K.~J.}\ \bibnamefont {Resch}}, \ and\
  \bibinfo {author} {\bibfnamefont {A.}~\bibnamefont {Zeilinger}},\ }\href
  {\doibase 10.1103/PhysRevLett.95.020403} {\bibfield  {journal} {\bibinfo
  {journal} {Phys. Rev. Lett.}\ }\textbf {\bibinfo {volume} {95}},\ \bibinfo
  {pages} {020403} (\bibinfo {year} {2005})}\BibitemShut {NoStop}%
\bibitem [{\citenamefont {Clauser}\ \emph {et~al.}(1969)\citenamefont
  {Clauser}, \citenamefont {Horne}, \citenamefont {Shimony},\ and\
  \citenamefont {Holt}}]{Clauser69}%
  \BibitemOpen
  \bibfield  {author} {\bibinfo {author} {\bibfnamefont {J.~F.}\ \bibnamefont
  {Clauser}}, \bibinfo {author} {\bibfnamefont {M.~A.}\ \bibnamefont {Horne}},
  \bibinfo {author} {\bibfnamefont {A.}~\bibnamefont {Shimony}}, \ and\
  \bibinfo {author} {\bibfnamefont {R.~A.}\ \bibnamefont {Holt}},\ }\href
  {\doibase 10.1103/PhysRevLett.23.880} {\bibfield  {journal} {\bibinfo
  {journal} {Phys. Rev. Lett.}\ }\textbf {\bibinfo {volume} {23}},\ \bibinfo
  {pages} {880} (\bibinfo {year} {1969})}\BibitemShut {NoStop}%
\bibitem [{\citenamefont {Bennett}\ \emph {et~al.}(1996)\citenamefont
  {Bennett}, \citenamefont {DiVincenzo}, \citenamefont {Smolin},\ and\
  \citenamefont {Wootters}}]{Bennett96}%
  \BibitemOpen
  \bibfield  {author} {\bibinfo {author} {\bibfnamefont {C.~H.}\ \bibnamefont
  {Bennett}}, \bibinfo {author} {\bibfnamefont {D.~P.}\ \bibnamefont
  {DiVincenzo}}, \bibinfo {author} {\bibfnamefont {J.~A.}\ \bibnamefont
  {Smolin}}, \ and\ \bibinfo {author} {\bibfnamefont {W.~K.}\ \bibnamefont
  {Wootters}},\ }\href {\doibase 10.1103/PhysRevA.54.3824} {\bibfield
  {journal} {\bibinfo  {journal} {Phys. Rev. A}\ }\textbf {\bibinfo {volume}
  {54}},\ \bibinfo {pages} {3824} (\bibinfo {year} {1996})}\BibitemShut
  {NoStop}%
\bibitem [{\citenamefont {Gurell}\ \emph {et~al.}(2007)\citenamefont {Gurell},
  \citenamefont {Bi\'emont}, \citenamefont {Blagoev}, \citenamefont {Fivet},
  \citenamefont {Lundin}, \citenamefont {Mannervik}, \citenamefont {Norlin},
  \citenamefont {Quinet}, \citenamefont {Rostohar}, \citenamefont {Royen},\
  and\ \citenamefont {Schef}}]{Gurell07}%
  \BibitemOpen
  \bibfield  {author} {\bibinfo {author} {\bibfnamefont {J.}~\bibnamefont
  {Gurell}}, \bibinfo {author} {\bibfnamefont {E.}~\bibnamefont {Bi\'emont}},
  \bibinfo {author} {\bibfnamefont {K.}~\bibnamefont {Blagoev}}, \bibinfo
  {author} {\bibfnamefont {V.}~\bibnamefont {Fivet}}, \bibinfo {author}
  {\bibfnamefont {P.}~\bibnamefont {Lundin}}, \bibinfo {author} {\bibfnamefont
  {S.}~\bibnamefont {Mannervik}}, \bibinfo {author} {\bibfnamefont {L.-O.}\
  \bibnamefont {Norlin}}, \bibinfo {author} {\bibfnamefont {P.}~\bibnamefont
  {Quinet}}, \bibinfo {author} {\bibfnamefont {D.}~\bibnamefont {Rostohar}},
  \bibinfo {author} {\bibfnamefont {P.}~\bibnamefont {Royen}}, \ and\ \bibinfo
  {author} {\bibfnamefont {P.}~\bibnamefont {Schef}},\ }\href {\doibase
  10.1103/PhysRevA.75.052506} {\bibfield  {journal} {\bibinfo  {journal} {Phys.
  Rev. A}\ }\textbf {\bibinfo {volume} {75}},\ \bibinfo {pages} {052506}
  (\bibinfo {year} {2007})}\BibitemShut {NoStop}%
\bibitem [{\citenamefont {Noel}\ \emph {et~al.}(2012)\citenamefont {Noel},
  \citenamefont {Dietrich}, \citenamefont {Kurz}, \citenamefont {Shu},
  \citenamefont {Wright},\ and\ \citenamefont {Blinov}}]{Noel12}%
  \BibitemOpen
  \bibfield  {author} {\bibinfo {author} {\bibfnamefont {T.}~\bibnamefont
  {Noel}}, \bibinfo {author} {\bibfnamefont {M.~R.}\ \bibnamefont {Dietrich}},
  \bibinfo {author} {\bibfnamefont {N.}~\bibnamefont {Kurz}}, \bibinfo {author}
  {\bibfnamefont {G.}~\bibnamefont {Shu}}, \bibinfo {author} {\bibfnamefont
  {J.}~\bibnamefont {Wright}}, \ and\ \bibinfo {author} {\bibfnamefont {B.~B.}\
  \bibnamefont {Blinov}},\ }\href {\doibase 10.1103/PhysRevA.85.023401}
  {\bibfield  {journal} {\bibinfo  {journal} {Phys. Rev. A}\ }\textbf {\bibinfo
  {volume} {85}},\ \bibinfo {pages} {023401} (\bibinfo {year}
  {2012})}\BibitemShut {NoStop}%
\bibitem [{\citenamefont {Wunderlich}\ \emph {et~al.}(2007)\citenamefont
  {Wunderlich}, \citenamefont {Hannemann}, \citenamefont {Körber},
  \citenamefont {Häffner}, \citenamefont {Roos}, \citenamefont {Hänsel},
  \citenamefont {Blatt},\ and\ \citenamefont {Schmidt-Kaler}}]{Wunderlich07}%
  \BibitemOpen
  \bibfield  {author} {\bibinfo {author} {\bibfnamefont {C.}~\bibnamefont
  {Wunderlich}}, \bibinfo {author} {\bibfnamefont {T.}~\bibnamefont
  {Hannemann}}, \bibinfo {author} {\bibfnamefont {T.}~\bibnamefont {Körber}},
  \bibinfo {author} {\bibfnamefont {H.}~\bibnamefont {Häffner}}, \bibinfo
  {author} {\bibfnamefont {C.}~\bibnamefont {Roos}}, \bibinfo {author}
  {\bibfnamefont {W.}~\bibnamefont {Hänsel}}, \bibinfo {author} {\bibfnamefont
  {R.}~\bibnamefont {Blatt}}, \ and\ \bibinfo {author} {\bibfnamefont
  {F.}~\bibnamefont {Schmidt-Kaler}},\ }\href {\doibase
  10.1080/09500340600741082} {\bibfield  {journal} {\bibinfo  {journal}
  {Journal of Modern Optics}\ }\textbf {\bibinfo {volume} {54}},\ \bibinfo
  {pages} {1541} (\bibinfo {year} {2007})}\BibitemShut {NoStop}%
\bibitem [{\citenamefont {Pinnington}\ \emph {et~al.}(1995)\citenamefont
  {Pinnington}, \citenamefont {Berends},\ and\ \citenamefont
  {Lumsden}}]{Pinnington95}%
  \BibitemOpen
  \bibfield  {author} {\bibinfo {author} {\bibfnamefont {E.~H.}\ \bibnamefont
  {Pinnington}}, \bibinfo {author} {\bibfnamefont {R.~W.}\ \bibnamefont
  {Berends}}, \ and\ \bibinfo {author} {\bibfnamefont {M.}~\bibnamefont
  {Lumsden}},\ }\href {\doibase 10.1088/0953-4075/28/11/009} {\bibfield
  {journal} {\bibinfo  {journal} {J. Phys. B.}\ }\textbf {\bibinfo {volume}
  {28}},\ \bibinfo {pages} {2095} (\bibinfo {year} {1995})}\BibitemShut
  {NoStop}%
\bibitem [{\citenamefont {Sackett}\ \emph {et~al.}(2000)\citenamefont
  {Sackett}, \citenamefont {Kielpinski}, \citenamefont {King}, \citenamefont
  {Langer}, \citenamefont {Meyer}, \citenamefont {Myatt}, \citenamefont {Rowe},
  \citenamefont {Turchette}, \citenamefont {Itano}, \citenamefont {Wineland},\
  and\ \citenamefont {Monroe}}]{Sackett00}%
  \BibitemOpen
  \bibfield  {author} {\bibinfo {author} {\bibfnamefont {C.~A.}\ \bibnamefont
  {Sackett}}, \bibinfo {author} {\bibfnamefont {D.}~\bibnamefont {Kielpinski}},
  \bibinfo {author} {\bibfnamefont {B.~E.}\ \bibnamefont {King}}, \bibinfo
  {author} {\bibfnamefont {C.}~\bibnamefont {Langer}}, \bibinfo {author}
  {\bibfnamefont {V.}~\bibnamefont {Meyer}}, \bibinfo {author} {\bibfnamefont
  {C.~J.}\ \bibnamefont {Myatt}}, \bibinfo {author} {\bibfnamefont
  {M.}~\bibnamefont {Rowe}}, \bibinfo {author} {\bibfnamefont {Q.~A.}\
  \bibnamefont {Turchette}}, \bibinfo {author} {\bibfnamefont {W.~M.}\
  \bibnamefont {Itano}}, \bibinfo {author} {\bibfnamefont {D.~J.}\ \bibnamefont
  {Wineland}}, \ and\ \bibinfo {author} {\bibfnamefont {C.}~\bibnamefont
  {Monroe}},\ }\href {\doibase 10.1038/35005011} {\bibfield  {journal}
  {\bibinfo  {journal} {Nature}\ }\textbf {\bibinfo {volume} {404}},\ \bibinfo
  {pages} {256} (\bibinfo {year} {2000})}\BibitemShut {NoStop}%
\bibitem [{\citenamefont {Chou}\ \emph {et~al.}(2013)\citenamefont {Chou},
  \citenamefont {Shu}, \citenamefont {Noel}, \citenamefont {Wright},
  \citenamefont {Graham},\ and\ \citenamefont {Blinov}}]{Chou13}%
  \BibitemOpen
  \bibfield  {author} {\bibinfo {author} {\bibfnamefont {C.-K.}\ \bibnamefont
  {Chou}}, \bibinfo {author} {\bibfnamefont {G.}~\bibnamefont {Shu}}, \bibinfo
  {author} {\bibfnamefont {T.}~\bibnamefont {Noel}}, \bibinfo {author}
  {\bibfnamefont {J.}~\bibnamefont {Wright}}, \bibinfo {author} {\bibfnamefont
  {R.}~\bibnamefont {Graham}}, \ and\ \bibinfo {author} {\bibfnamefont
  {B.}~\bibnamefont {Blinov}},\ }\href@noop {} {\bibfield  {journal} {\bibinfo
  {journal} {Bull. Am. Phys. Soc.}\ }\textbf {\bibinfo {volume} {58}},\
  \bibinfo {pages} {157} (\bibinfo {year} {2013})}\BibitemShut {NoStop}%
\end{thebibliography}
\end{document}